\documentclass{article}
\usepackage[dvips]{graphicx}
\newcommand{\be}{\begin{equation}}
\newcommand{\en}{\end{equation}}
\newcommand{\bea}{ \begin{eqnarray}}
\newcommand{\ena}{\end{eqnarray} }
\newcommand{\grs}{\biggl\{}
\newcommand{\grd}{\biggr\}}
\newcommand{\ii}{{\bf i}}
\newcommand{\jj}{{\bf j}}

\newcommand{\rr}{{\bf r}}
\newcommand{\es}{{\bf s}}
\newcommand{\oo}{{\bf 0}}
\newcommand{\pp}{{\bf p}}
\newcommand{\qq}{{\bf q}}
\newcommand{\intphi}{\int \prod_{\ii} D
 \varphi_{\ii} \: \:}
\newcommand{\intau}{\int_{0}^{\beta} d\tau }
\newcommand{\intpsi}{\int \prod_{\ii}D^2  \psi_{\ii} \: \:}
\newcommand{\sumij}{\sum_{\ii \jj}}
\newcommand{\sumi}{\sum_{\ii}}

\newcommand{\psii}{\psi_{\ii}}
\newcommand{\psij}{\psi_{\jj}}
\newcommand{\varphii}{\varphi_{\ii}}
\newcommand{\varphij}{\varphi_{\jj}}
\newcommand{\varphir}{\varphi_{\rr}}
\newcommand{\varphis}{\varphi_{\es}}
\newcommand{\psir}{\psi_{\rr}}
\newcommand{\psis}{\psi_{\es}}
\newcommand{\phii}{\phi_{\ii}}
\newcommand{\phij}{\phi_{\jj}}

\begin{document}
\begin{center}
{\huge \bf Mean Field Theory for Josephson Junction Arrays with
Charge Frustration}\\
\vskip 0.3 truein
{\bf  G. Grignani, A. Mattoni, P. Sodano and A. Trombettoni
\footnote{Present
address: S.I.S.S.A., Via Beirut 4, 34814 Trieste, Italy}}
\vskip 0.3truein
 {\it Dipartimento di Fisica, 
Universit\`a di Perugia, Via A. Pascoli I-06123 Perugia,
Italy}\\
\vskip 0.3truein
DFUPG-198-99
\vskip 0.3truein
{\bf Abstract}
\vskip 0.1truein
\end{center}
Using the path integral approach, we provide an explicit
derivation of the equation for the phase boundary for quantum
Josephson junction arrays with offset charges and non-diagonal
capacitance matrix. 
For the model with nearest neighbor capacitance
matrix and uniform offset charge $q/2e=1/2$, we determine, 
in the low critical temperature expansion, the most relevant contributions 
to the equation for the phase boundary.
We explicitly construct the charge distributions
on the lattice corresponding to the lowest energies.
We find a reentrant behavior even with a short ranged interaction.

A merit of the path integral approach  is that it allows to 
provide an elegant derivation of the Ginzburg-Landau free energy for 
a general model with charge frustration and non-diagonal capacitance
matrix. The partition function factorizes as a product of a 
topological term, depending only on a set of integers, 
and a non-topological one, which is explicitly evaluated.

\vskip 0.5truein
\section{Introduction}
Josephson junction arrays (JJA) and granular superconductors,
namely systems of metallic grains embedded in an insulator, become 
superconducting in two steps~\cite{sima}. First, 
at the bulk critical temperature
each grain develops a superconducting gap but the phases
of the order parameter on different grains are uncorrelated.
Then, at a lower temperature $T_c$,
the Cooper pair tunneling between grains gives rise to a long-range 
phase coherence and the system as a whole exhibits a phase transition to 
a superconducting state. 

The phase transition is governed by the competition between 
the Josephson tunneling, characterized by a Josephson coupling 
energy $E_J$~\cite{Joseph}, and the 
Coulomb interaction between Cooper pairs,
described by a charging energy $E_C$~\cite{And,Ab}.
In classical junction arrays the Josephson coupling $E_J$ is dominant,
the transition separates a superconducting low temperature phase
from a normal high temperature phase.
When $E_C$ is comparable to $E_J$ (small grains)
charging effects give rise to a quantum dynamics.
The grain capacitance is small, 
so that the energy cost of Cooper pair tunneling may be higher then the 
energy gained by the formation of a phase-coherent state.
Zero point fluctuations of the phase
may destroy the long range superconducting order 
even at zero temperature (see for example~\cite{sima}).

Within the framework of the BCS theory, Efetov~\cite{Ef} derived an 
effective quantum Hamiltonian in terms of the phases $\varphi_{\ii}$
of the superconducting order parameter at the grain $\ii$,
and their conjugate variables ${n}_{\ii}$ number of Cooper pairs.
Efetov's Hamiltonian for the quantum phase model reads
\be \label{1}
{H}=\frac{1}{2}\sum_{\ii \jj}C_{\ii \jj}^{-1}{Q}_{\ii}
{Q}_{\jj}-E_J\sum_{<\ii \jj>}\cos({\varphi_{\ii}}-
{\varphi_{\jj}})
\en
$$
{Q}_{\ii}=2e {n}_{\ii} \hspace{1cm} [{\varphi_{\ii}},
{n}_{\ii}]=\;i\;\delta_{\ii \jj}\ ,
$$
where ${Q}_{\ii}$ is the excess of charge due to Cooper pairs
(charge $2e$) on the site $\ii$ of a square lattice in D-space dimension and
$C_{\ii \jj}$ is the capacitance matrix describing the
electrostatic coupling between Cooper pairs.
The diagonal elements of the inverse matrix $C_{\ii\jj}^{-1}$ provide the
charging energy: $E_C=e^2 C_{\ii\ii}^{-1}/2\equiv e^2 /2 C_0$, where $C_0$ is
the self-capacitance.

The superconductor-insulator transition depends crucially on the 
spatial dimensionality $D$. For $D=1$ there may exist also other 
phases~\cite{larkin}. For $D=2$ the system exhibits a richly 
structured phase diagram (see for example~\cite{FaSh1,FaSh2}).
In higher dimensions it is believed that the mean field theory 
analysis provides qualitatively correct results.

It is relevant to understand how the transition from insulator to
superconductor depends on the relevant constitutive parameters - such as the 
capacitances of, and between, the junctions - as well as on 
external parameters 
such as the temperature, offset charges and external magnetic fields.

Much work has been done to study the phase diagram of quantum JJA,
in the $T/E_C$-$E_C/E_J$ plane~\cite{sima}. The analysis uses mean field 
theory~\cite{Ef,Sima1,Sima2,Sima3,Fa,Kiss,Simkin,GC} as well as the 
renormalization group approach~\cite{Jo,Fa1}.
There is the claim that the phase diagram -under suitable circumstances- 
may exhibit a reentrant character with the superconducting phase 
existing between 
upper and lower critical temperature~\cite{Sima1,Sima2,Simkin}. 
In~\cite{Sima1,FiSt1} the influence 
of the Coulomb energy on the transition temperature was investigated for a 
model with a diagonal capacitance matrix. The effects of off-diagonal 
terms in the charging energy were investigated by several authors 
within the mean field theory 
approximation~\cite{Ef,Sima3,Fa,Fa1,Don,FiSt2,Fi}:
while it is widely believed that the nearest neighbors interaction enhance 
the transition temperature $T_c$ by lowering the energy cost for a Cooper 
pair to tunnel from one neutral grain to another~\cite{Fa}, there is 
still some 
dispute on whether there is a reentrance or not for models with non-diagonal
capacitance matrix~\cite{Fa,FiSt2,Fi}.

It is relevant for physical applications to consider the effect
of a background of external charges on the 
superconductor-insulator transition 
of a quantum JJA~\cite{FaSh1,FaSh2,RS,Luc}. Such an offset of charges
might arise in physical systems as a result of charged impurities or by 
application of a gate voltage between the array and the ground.
In the former case offset charges are distributed randomly on 
the lattice while in the latter case they play the role of a 
chemical potential for charges.
They might be regarded as effective charges $q_{\ii}$ on the 
sites of the lattice. When $q_{\ii}\ne 2e$ the offset charges 
cannot be eliminated by Cooper pair tunneling.

Offset charges frustrate the attempts of the system to minimize 
the energy of the charge distribution of the ground state.
Consequently the charging energy of any excitations is smaller compared
to the unfrustrated case and superconductivity is enhanced.
With offset charges the Hamiltonian (\ref{1}) becomes
\be \label{7}
{H} = \frac{1}{2}\sum_{\ii \: \jj} 
C_{\ii \:\jj}^{-1} ({Q}_{\ii}+q_{\ii})( 
{Q}_{\jj}+q_{\jj})-E_{J} \sum_{< \ii \: \jj >}
\cos ({\varphi}_{\ii}-{\varphi}_{\jj}).
\en

In order to study the effect of charge frustration
on the phase diagram of the system described by the Hamiltonian (\ref{7}), 
it is our purpose to revisit the mean field theory of quantum JJA using the 
path-integral method.
The approach uses the Hubbard-Stratonovich~\cite{hub} 
representation for the partition function in terms of coarse-grained 
classical local variables $\psi_{\ii}$ for which the effective action 
is computed~\cite{Don}.
We find a reentrant behavior for models with a nearest neighbor 
capacitance matrix and a uniform offset charge $q_{\ii}=e$, even if the 
interaction among grains is short ranged.
We find analytically the equation which determines  
the critical temperature as a function of $E_J/E_C$.
This allows us to analyze the low temperature limit of the theory 
and to find the regimes in which a reentrant behavior might be observed.

In section 2 we review the self consistent mean field theory 
approximation within the Hamiltonian formalism
for quantum JJA.
We study the eigenvalue equation
of the mean field Hamiltonian with diagonal capacitance,
and uniform offset charge $q_{\ii}=e$ showing explicitly that
at zero temperature there is superconductivity for all values of 
$\alpha=zE_J/4E_C$.

In section 3 we use the coarse grained approach to compute  
the Ginzburg-Landau free energy for quantum JJA with charge 
frustration and a general Coulomb interaction matrix.
The path integral
providing the phase correlator 
needed to investigate the critical behavior of the system, is explicitly 
computed.

In section 4, from the Ginzburg-Landau free energy, we
derive, within the mean field theory approximation,
the analytical form of the critical line equation. 
The phase diagram  is drawn in the diagonal case for a 
generic external charge distribution.
We then analyze the low temperature limit of a system 
with nearest neighbor interaction matrix and find a reentrant 
behavior when a uniform background of external charges
$q_{\ii}=e$ is considered.

Section 5 is devoted to some concluding remarks.
The appendices contain the derivation of some relevant formulas
of the main text.

\section{Self-consistent mean field theory in the Hamiltonian approach}

Mean field theory for quantum JJA with diagonal capacitance matrix
was first used by Symanek \cite{Sima1}. The approximation 
consists in replacing the Josephson coupling of the phase on a given 
island $\ii$ to its neighbors by an average coupling so that:
\be \label{2}
E_J\sum_{<\ii \jj>}\cos({\varphi_{\ii}}-{\varphi_{\jj}})=
z E_J<\cos {\varphi}> \sum_{\ii} \cos {\varphi_{\ii}}.
\en 
In (\ref{2}) $z$ is the coordination number; it is assumed also 
that $<\cos{\varphi}>$ does not depend on the island index 
$\ii$ and the choice  $<\sin {\varphi_{\ii}}>=0$, which provides a real
order parameter, has been made.

Within the  mean field approximation the Hamiltonian 
(\ref{1}) becomes
\be \label{3}
{H}_{mf}=\frac{1}{2}\sum_{\ii \jj}C_{\ii \jj}^{-1}
{Q_{\ii}}{Q_{\jj}}-z E_J<\cos {\varphi}> 
\sum_{\ii}\cos {\varphi_{\ii}}
\en
and the order parameter  
$<\cos {\varphi}>$ is evaluated self-consistently from (\ref{3}).
For a diagonal capacitance matrix ($C_{\ii \jj}=C_0
\delta_{\ii \jj}$) mean field theory computation are 
very simple since (\ref{3}) describes on each site a 
quantum particle in a periodic potential 
$\cos {\varphi}_{\ii}$~\cite{Sima1}.

The eigenfunction of the array is a product of 
eigenfunctions $\psi_n(\varphi)$ describing the 
individual islands and satisfying the Mathieu equation~\cite{Abramowitz}
\be \label{4}
\bigg(-\frac{d^2}{d\varphi^2}-
\frac{z E_J}{4E_{C}}<\cos \varphi> \cos \varphi \bigg)
\psi_n (\varphi)=\frac {E_n}{4 E_C} 
\psi_n (\varphi)
\en
with periodic boundary conditions $\psi_n (\varphi)=
\psi_n (\varphi+2\pi)$.

 It is well known that the Mathieu equation 
admits also antiperiodic solutions, 
$\psi_n(\varphi)=-\psi_n(\varphi+2\pi)$ (see appendix A). 
If both periodic and antiperiodic solutions are used,
the general solution of (\ref{4}) does not have a definite periodicity 
and, consequently, the charges ${n}_{\ii}$ take continuous eigenvalues.
Such continuous eigenvalues are expected to be relevant in the description
of continuous flows of currents due for example to ohmic shunt 
resistances~\cite{schon,likh}.
Although the superposition of both periodic and antiperiodic solutions
yields to a reentrant behavior
even in the unfrustrated dissipationless
diagonal model~\cite{Sima2,Sima3,Sima4}, this superposition is not
allowed in describing physical situations in which the only
excitations are Cooper pairs of charge $2e$~\cite{Ef,Fa,FaSh1}.
Thus the use of both periodic and antiperiodic solutions does 
not have physical significance in the models considered in this paper.

The mean field self-consistency condition gives
\be \label{5}
<\cos \varphi>=\frac{\sum_n e^{-\beta E_n}
<\psi_n|\cos \varphi |\psi_n>}{\sum_n e^{-\beta E_n}}
\en  
with $\beta=1/k_B T$. 
For high temperatures or low $E_J$ only the 
solution $<\cos \varphi>=0$ exists and there is not superconductivity.
For low temperatures or high $E_J$  $<\cos \varphi>\ne 0$ and the system
as a whole behaves as a superconductor.

From (\ref{5}) one gets
the equation for the critical temperature \cite{Sima1} 
\be 
\label{6}
\alpha=\frac{\sum_{n=-\infty}^{+\infty}
e^{-\frac{4}{y}n^2}}{  
\sum_{n=-\infty}^{+\infty}\frac{1}{1-4n^2}e^{-\frac{4}{y}n^2} } 
\en
with  $y=k_B/T_c E_C$ and $\alpha=z E_J/4E_C$.

In fig.1 we plot $T_c$ as a function of $\alpha$.
\begin{figure}[ht]
\begin{center}
\includegraphics[width=80mm,height=8cm]{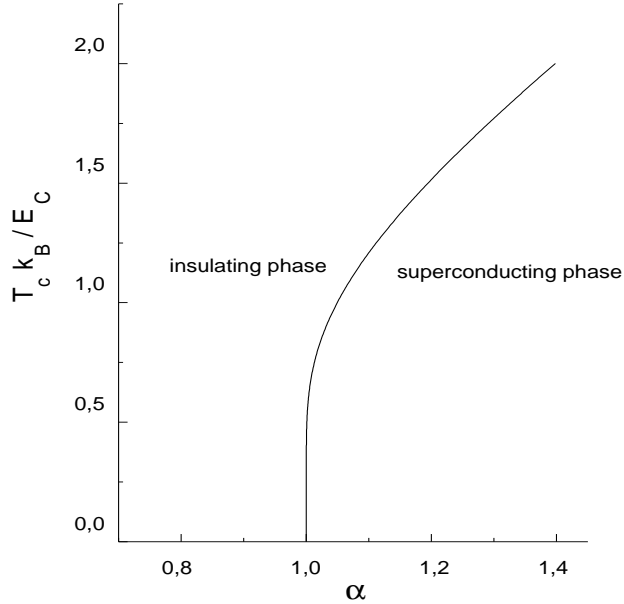}
\caption{Phase diagram for the diagonal model without charge 
frustration.}
\end{center}
\end{figure}

If one considers  a 
diagonal capacitance matrix and uniform offset charges of 
magnitude $e$ on each site ($q_{\ii}/2 e=1/2)$,
the Hamiltonian reads
\be \label{7d}
{H_d} = \frac{1}{2 C_0}\sum_{\ii} 
({Q}_{\ii}+q_{\ii})( 
{Q}_{\ii}+q_{\ii})-E_{J} \sum_{< \ii \: \jj >}
\cos ({\varphi}_{\ii}-{\varphi}_{\jj})\ .
\en
Mean field theory of this model leads to a Schr\"odinger equation 
of the form
\be \label{8d}
\Big[
-\frac{d^2}{d\varphi^2}-2i \frac{q}{2 e} \frac{d}{d\varphi}+
\bigg(\frac{q}{2 e} \bigg)^2-\alpha <\cos \varphi> \cos \varphi 
\Big]\psi_n(\varphi)=\frac{E_n}{4 E_C} \psi_n(\varphi).
\en
Redefining the phase of the wave function as 
$$
\psi_n(\varphi)=e^{-i \frac{q}{2e} \varphi} \rho_n(\varphi)
$$
(\ref{8d}) reduces to a Mathieu equation for $\rho_n(\varphi)$
\be \label{9}
\frac{d^2\rho_n}{d\varphi^2}+\bigg(\frac{\lambda}{4}-\frac{v}{2}
\cos \varphi \bigg)\rho_n =0
\en
with $\lambda_n=E_n /E_C$ and $v=- z E_J <\cos \varphi>/2 E_C$.
Equations (\ref{7d}),(\ref{8d}),(\ref{9}) lead to the following
modification of (\ref{6}) [see Appendix A]:
\be \label{10}
\alpha=\frac{e^{-\frac{1}{y}}+\sum_{n=1}^{+\infty}
e^{-\frac{4}{y}(n+\frac{1}{2})^2}}{\frac{4+y}{4y}e^{-\frac{1}{y}}
+\sum_{n=1}^{+\infty}\frac{1}{1-4(n+\frac{1}{2})^2}
e^{-\frac{4}{y}(n+\frac{1}{2})^2} }
\en
which - at variance with the unfrustrated model - 
exhibits superconductivity even for infinitesimal values of $\alpha$.
This feature is shown in fig.2 which also shows the absence 
of a reentrant behavior at low T.
\begin{figure}[ht] 
\begin{center}
\includegraphics[width=8cm,height=8cm]{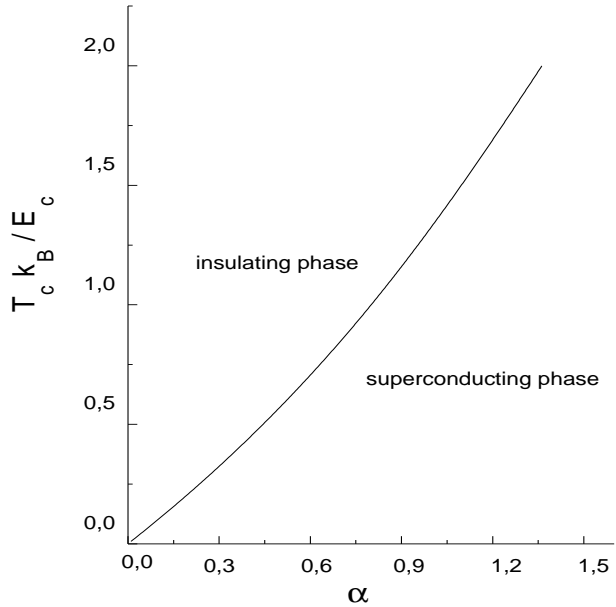}
\end{center}
\caption{Phase diagram of the diagonal model with half-integer 
charge frustration.}   
\end{figure}

For frustrated models with non diagonal capacitance 
matrix, the self-consistent mean-field theory approximation becomes very 
cumbersome and one should resort to the more powerful functional approach.
A reentrance at a low $T_c$ is expected
at least when the interaction between grains is long-ranged~\cite{RS,FiSt1}.

\section{Ginzburg-Landau free energy}
The partition function for the frustrated off-diagonal model
is given by
\be
\label{msz}
Z={\rm Tr} e^{-\beta H}=\sum_n \left<\psi_n\left|e^{-\beta H}
\right|\psi_n\right>
\en 
where $H$ is given in (\ref{7}) and the sum is extended only to 
states of charge $2e$ and thus with definite periodicity.

In the functional approach $Z$ reads
\be 
Z=\int \prod_{\ii} D\varphi_{\ii} \exp \bigg\{ - 
\int_0^{\beta}d\tau L_E\left(\varphi_{\ii}(\tau),
\frac{d\varphi_{\ii}}{d\tau}(\tau)\right) \bigg\}
\en  
where the Euclidean Lagrangian $L_E$  can be derived from
\be \label{11}
L=\frac{1}{2}\bigg(\frac{\hbar}{2e} \bigg)^2\sum_{\ii \jj}C_{\ii \jj}
\frac{d\varphi_{\ii}}{dt}   \frac{d\varphi_{\jj}}{dt}-\bigg(\frac{\hbar}{2e} 
\bigg)\sum_{\ii}\frac{d\varphi_{\ii}}{dt}q_{\ii} +E_J \sum_{<\ii\jj>}
\cos (\varphi_{\ii}-\varphi_{\jj})
\en
by replacing $i t / \hbar \to \tau$.
The path integral that one should compute is then given by:  
\be  \label{12}
Z= \intphi \exp  \bigg\{ \int_{0}^{\beta} d\tau    
\big[ - \frac{1}{2} \sum_{\ii \jj} C_{\ii \jj}
\frac{\dot{\varphi_{\ii}}}{2e} \frac{\dot{\varphi_{\jj}}}{2e} 
+ i \sum_{\ii}q_{\ii} \frac{\dot{\varphi_{\ii}}}{2e}+ 
\frac{E_J}{2}\sum_{ \ii \: \jj }e^{ i\varphi_{\ii}}
\gamma_{\ii \jj} e^{-i\varphi_{\jj}} \big]\bigg\}
\en
where $-\infty < \varphi_{\ii} < + \infty $, 
$\varphi_{\ii}(0)=\varphi_{\ii}(\beta)+2\pi n_{\ii}$
and $\gamma_{\ii \jj}=1$ if $\ii,\jj$ are nearest 
neighbors and equals zero otherwise.
The integers $n_{\ii}$ appearing in this
boundary conditions take into account the $2\pi$-periodicity of the
states $\psi_n$ appearing in (\ref{msz}).

In order to derive the Ginzburg-Landau free energy for the order parameter,
it is convenient to carry out the integration over the phase 
variables by means of the Hubbard-Stratonovich 
procedure~\cite{hub}: using the identity
\be \label{gauss}
 e^{J^{+} \Gamma J} =
\frac{\det \Gamma^{-1}}{\pi^{N}} \intpsi
e^{- \psi^{+} \Gamma^{-1} \psi -J^{+} \psi - \psi^{+} J} 
\en
the partition function may be rewritten as 
\be \label{13}
 Z = \int \prod_{\ii} D\psi_{\ii} D\psi_{\ii}^*  e^{
\intau (-\frac{2}{E_J} \sumij \psii^*
\gamma_{\ii \jj}^{-1} \psij )} e^{-S_{Eff}[\psi]}.
\en
where the effective action for the auxiliary Hubbard-Stratonovich
field $\psi_\ii$, $S_{Eff}[\psi]$, is given by 
\be \label{14}
S_{Eff}[\psi]= - \log  \grs \intphi \exp \bigg\{ \intau [
-\frac{1}{2}\sumij C_{\ii \jj}\frac{\dot{\varphii}}{2e} 
\frac{\dot{\varphij}}{2e} +$$ $$  +i \sumi ( q_{\ii} 
\frac{\dot{\varphii}}{2e}-\psii e^{i\varphii}-\psii^* 
e^{-i \varphii})] \bigg\} \grd .  
\en
The Hubbard-Stratonovich field $\psi_{\ii}$ may be regarded as the order 
parameter for the insulator-superconductor phase transition since it
turns out to be proportional to 
$<e^{i\varphi_{\ii}}>$, as it can be easily seen from the classical 
equations of motion.
From (\ref{14}) the Ginzburg-Landau free-energy may be derived
by integrating out the phase field $\phi_{\ii}$.

Since the phase transition is second order~\cite{Simkin1},
close to the onset of superconductivity, the order parameter 
$\psi_{\ii}$ is small. One may then expand the effective action up to the 
second order in $\psi_{\ii}$, getting
\be \label{15}
S_{Eff}[\psi]= S_{Eff}[0]+\intau \intau'
G_{\rr \es}(\tau,\tau')\psir(\tau) \psis^*(\tau')+ \cdots
\en
where $G_{\rr \es}$ is the phase correlator
\be \label{16}
G_{\rr \es}(\tau, \tau')= \frac{
\delta^2S_{Eff}[\psi]}{\delta \psir(\tau) \delta \psis (\tau')} 
\Bigg|_{\psi,\psi^* = 0} =\langle e^{i\varphir(\tau)
-i \varphis(\tau')} \rangle_0. 
\en
The partition function (\ref{13}), can be written as 
\be  \label{17}
Z =\int \prod_{\ii} d\psi_{\ii} d\psi_{\ii}^* e^{-F[\psi]}
\en  
where $F[\psi]$ is the Ginzburg-Landau free energy; 
due to (\ref{14},\ref{15}), up to the second order in $\psi_\ii$, one has
\be \label{18}
 F[\psi]=\intau \intau'  \sumij
\psii^*(\tau) [ \gamma_{\ii \jj}^{-1} \delta(\tau-\tau')-
G_{\ii \jj}(\tau,\tau')]\psij(\tau').  
\en

We shall now compute the phase correlator
$G_{\rr \es}$ by evaluating  the expectation value in (\ref{17})
by means of the path integral over the phase variables
$\varphi_{\ii}(\tau)$. 
In performing this integration
one should take into account that the field 
configurations satisfy
\be \label{19}
\varphi_{\ii} (\beta)-\varphi_{\ii} (0)=2\pi n_{\ii}\ .
\en
For this purpose it turns out very convenient to 
untwist the boundary conditions by decomposing the phase field in terms 
of a periodic field $\phi_{\ii}(\tau)$ and a term linear in 
$\tau$ which takes into account the boundary conditions (\ref{19});
namely, one sets
\be \label{20}
\varphi_{\ii}(\tau)=\phi_{\ii}(\tau)+\frac{2\pi}{\beta}n_{\ii}\tau\ ,
\en
with $\phi_{\ii}(\beta)=\phi_{\ii}(0)$.
Summing over all the phases $\varphi_{\ii}(\tau)$ amounts then to 
integrate over the periodic field $\phi_{\ii}$ and to sum over the 
integers $n_{\ii}$.
As a result the phase correlator factorizes as the product of a 
topological term depending on the integers $n_{\ii}$ and a 
non-topological one; namely
$$
G_{\bf r s}(\tau;\tau') =
\frac{\int D \phi_{\ii} e^{i\phi_{\bf r}(\tau)-
i\phi {\bf_s}(\tau')} \exp \grs \intau (-\frac{1}{2}C_{\bf ij}
\dot{\frac{\phii}{2e}}\dot{\frac{\phij}{2e}})\grd}  
{\int D \phi_{\ii} \exp \grs \intau (-\frac{1}{2}C_{\bf ij}
\dot{\frac{\phii}{2e}}\dot{\frac{\phij}{2e}})\grd} \cdot $$
\be \label{21}
\cdot \> \frac{  
\sum_{[n_{\ii}]} 
e^{i\frac{2 \pi}{\beta}(n_{\rr}\tau-n_{\es}\tau ')} 
e^{\grs \sum_{\bf i j}-\frac{\pi^2}{2 \beta e^2} C_{\bf ij}
n_{\ii}n_{\jj}+ \sum_{\ii}2i \pi   \frac{q_{\ii}}{2e}n_{\ii} \grd } }
{  \sum_{[n_{\ii}]} e^{\grs \sum_{\bf i j}-\frac{\pi^2}{2 \beta e^2}C_{\bf ij}
n_{\ii}n_{\jj}+\sum_{\ii}2i\pi \beta
\frac{q_{\ii}}{2e}n_{\ii} \grd }}\ . 
\label{corr1}
\en
After a lengthy computation, the first (non-topological) factor 
appearing in the l.h.s. 
of equation (\ref{corr1}) has the following simple expression
[see Appendix B]:
\be \label{22}
\delta_{\rr \es} \exp \grs -2e^2C^{-1}_{\rr \rr}\biggl(|\tau-\tau'|-
\frac{(\tau-\tau')^2}{\beta} \biggr)\grd . 
\en  
The sum over the integers in the topological factor in (\ref{corr1}) is
done by means of the Poisson resummation 
formula
$$
|\det G|^{\frac{1}{2}} \sum_{[n_{\ii}]} e^{ - \pi (n-a)_{\ii}G_{\ii \jj}
(n-a)_{\jj}}=
\sum_{[m_{\ii}]} e^{ - \pi m_{\ii}(G^{-1})_{\ii\jj}m_{\jj}
-2\pi i m_{\ii}a_{\ii}} \label{PR}\ .
$$
Thus eq.(\ref{corr1}) becomes
$$
G_{\rr \es}(\tau,\tau')=\delta_{\rr \es}e^{ -2e^2C^{-1}_{\rr \rr} 
|\tau-\tau'|} \>\> \cdot $$ 
\be \label{corr2}
\cdot \> \frac{\sum_{[n_{\ii}]}e^{  -\sum_{\ii \jj} 2e^2\beta 
C_{\ii \jj}^{-1}(n_{\ii}+\frac{q_{\ii}}{2e})(n_{\jj}+
\frac{q_{\jj}}{2e})-\sum_{\ii}4e^2C_{\rr \ii}^{-1}(n_{\ii}+
\frac{q_{\ii}}{2e})(\tau-\tau')}  }{\sum_{[n_{\ii}]}e^{\sum_{\ii \jj} 2
\beta e^2C_{\ii \jj}^{-1}(n_{\ii}+\frac{q_{\ii}}{2e})(n_{\jj}+
\frac{q_{\jj}}{2e}) }  } 
\en
with $n_{\ii}$ assuming all integer values and  
$\sum_{[n_{\ii}]}$ being a sum over all the configurations. 

By means of a Euclidean-time Fourier transform, the fields $\psi_{\ii}$
are written as
$$
\psi_{\ii}(\tau)=\frac{1}{\beta}\sum_{\mu}
\psi_{\ii}(\omega_{\mu})e^{i\omega_{\mu}\tau}\ ,
$$
where $\omega_\mu$ are the Matsubara frequencies.
As a consequence, the phase correlator $G_{\ii\jj}$ can be expressed 
as 
\be \label{24}
G_{\ii\jj}(\tau;\tau')=\frac{1}{\beta}\sum_{\mu \mu'}
G_{\ii\jj}(\omega_{\mu} ;\omega_{\mu'})e^{i\omega_{\mu}\tau}
e^{i\omega_{\mu'}\tau'}\ .
\en
 From (\ref{corr2}) one can show that 
$G_{\rr \es}(\omega_{\mu};\omega_{\mu}')$
is diagonal in the Matsubara frequencies and can be written as
\be \label{25}
G_{\rr \es}(\omega_{\mu};\omega_{\mu}')=
G_{\rr }(\omega_{\mu})\cdot
\delta_{\rr \es}\cdot\delta(\omega_{\mu}+\omega_{\mu'})
\en
with
\be \label{26}
G_{\rr}(\omega_{\mu})=\frac{1}{2E_{c}}\cdot  
\sum_{[n_{\ii}]}  \frac{  \;  e^{-\frac{4}{y} 
\sum_{\ii \jj}\frac{U_{\ii \jj}}{U_{\oo\oo}}  
(n_{\ii}+\frac{q_{\ii}}{2e})(n_{\jj}+\frac{q_{\jj}}{2e})}}
{ 1-4[\sum_{\jj}\frac{U_{\bf r j}}{U_{\oo \oo}} (n_{\ii}+
\frac{q_{\ii}}{2e})-i \omega_{\mu}]^2}\cdot\frac{1}{Z_0}\ .
\en
In (\ref{27}) $Z_0$ is given by 
$$ Z_0=  \sum_{[n_{\ii}]}e^{-\frac{4}{y} \sum_{\ii 
\jj}\frac{U_{\ii \jj}}{U_{\oo \oo}}  
(n_{\ii}+\frac{q_{\ii}}{2e})(n_{\jj}+\frac{q_{\ii}}{2e})}\ .  
$$
with
$U_{\ii \jj}=C_{\ii \jj}^{-1}$, $ E_C=e^2C_{\rr \rr}^{-1}/2$ and 
$y=k_B T_c/E_C$.
In terms of Matsubara frequencies the Ginzburg-Landau 
free energy (\ref{18}) becomes
\be \label{27}
F[\psi]=\frac{1}{\beta} \sum_{\mu  \ii \jj}
\psi_{\ii}^*(\omega_{\mu})
\biggl[ \frac{2}{E_J}\gamma_{\ii \jj}^{-1}-G_{\ii}
(\omega_{\mu})\delta_{\ii \jj} \biggr]\psi_{\jj}(\omega_{\mu})\ .
\en
This is our starting point for any analysis of the phase 
boundary between the insulating and the superconducting phases in 
JJA with arbitrary capacitance matrix and with
charge frustration.

\section{Mean field theory analysis}
In the following we shall derive the equation
determining the phase boundary in the plane $(\alpha, K_B T_c/E_C)$,
in mean field theory and for a system with arbitrary capacitance 
matrix and a uniform distribution of off-set charges. 
For this purpose it is convenient to expand the fields 
$\psi_{\ii}(\omega_\mu)$ and $G_{\ii}(\omega_\mu)$ 
in terms of the vectors of the 
reciprocal lattice $\qq$. One has
\begin{eqnarray} 
\psi_{\ii}(\omega_{\mu})&=&\frac{1}{N}\sum_{\qq}
\psi_{\qq}(\omega_{\mu})e^{i\qq\cdot \ii}\label{28}\\
G_{\ii}(\omega_{\mu})&=&\frac{1}{N}\sum_{\qq}
G_{\qq}(\omega_{\mu})e^{i\qq\cdot \ii}\label{giq}\ .
\end{eqnarray}
Moreover
\be \label{29}
\gamma_{\ii \jj}^{-1}=\frac{1}{N}\sum_{\qq} 
\gamma_{\qq}^{-1}e^{i\qq\cdot( \ii-\jj)}.
\en
where $\gamma_{\qq}^{-1}$ is the inverse of  the Fourier transform 
of the Josephson coupling strength $\gamma_{\ii\jj}$ which equals 1 
for $\ii,\jj$ nearest neighbors and 0 otherwise. As a consequence
$$
\gamma_{\qq}^{-1}=\frac{1}{\sum_{\pp}e^{-i \qq\cdot \pp}}
$$
where $\pp$ is a vector connecting two nearest neighbors sites.
Expanding in $\qq$ one gets
\be\label{gammaq}
\gamma_{\qq}^{-1}=\frac{1}{z}+\frac{{\qq}^2 a^2}{ z^2}+\cdots
\en
where $a$ is the lattice spacing and $z$ the coordination number.
The first term in (\ref{gammaq}) provides the mean field theory 
approximation which, as expected, is exact in the limit of large 
coordination number.

The Ginzburg-Landau free energy  (\ref{27}), reads
$$
F[\psi]=
\frac{1}{\beta N}\sum_{\mu \qq\qq'}\psi_{\qq}(\omega_{\mu})^*
[\gamma_{\qq}^{-1}\delta_{\qq\qq'}-\frac{G_{\qq-\qq'}
(\omega_{\mu})}{N}]\psi_{\qq'}(\omega_{\mu})
\simeq
$$
Using (\ref{gammaq}) and keeping only terms of zero-th order in $\omega_\mu$
and $\qq$ one obtains the mean field theory 
approximation to the coefficient of the quadratic term of $F$ 
\be \label{30}
\simeq \frac{1}{\beta N}\sum_{\qq \mu}
\bigg[ \frac{2}{E_J z}-G_{\bf 0}(0)
+\cdots \bigg] |\psi_{\qq}(\omega_{\mu})|^2\ .
\en
The equation for the phase boundary line then reads as
\be \label{eqgenlin} 
1=z\frac{E_J}{2}G_{\bf 0}(0)
\en
with
\be \label{Gconr}
G_{\bf 0}(0)=\frac{1}{N}\sum_{\rr}G_{\rr}(0).
\en
Equation (\ref{eqgenlin}) determines
the relation between $T_c$ and $\alpha$ at the phase boundary.

For a uniform distribution of offset charges eq.(\ref{eqgenlin}) 
simplifies further since in (\ref{Gconr}) $G_{\rr}$ does not depend on $\rr$.
As a consequence, the phase boundary equation becomes
\be \label{lineacritica}
1=\alpha\cdot   \sum_{[n_{\ii}]}  \frac{  \;  e^{-\frac{4}{y} \sum_{\ii \jj} 
\frac{U_{\ii \jj}}{U_{\oo\oo}}  (n_{\ii}+\frac{q}{2e})(n_{\jj}+
\frac{q}{2e})}} { 1-4[\sum_{\jj}\frac{U_{\bf 0 j}}{U_{\bf 00}} 
(n_{\ii}+\frac{q}{2e})]^2}
\cdot\frac{1}{Z_0}   
\en
with
$$ 
\alpha=\frac{zE_J}{4E_{c}}
\hspace{0.5cm}\mbox{and}\hspace{0.5cm}
Z_0=  \sum_{[n_{\ii}]}e^{-\frac{4}{y} \sum_{\ii \jj}
\frac{U_{\ii \jj}}{U_{\bf 00}}(n_{\ii}+\frac{q}{2e})(n_{\jj}+\frac{q}{2e})}.  
$$

In the following we shall derive the physical implications of 
(\ref{lineacritica}) in a variety of models describing JJA.

\subsection{Self-charging model}
For a diagonal capacitance matrix, $U_{\ii \jj}= \delta_{\ii \jj}U_{\oo}$, 
one singles out only the self-interaction of plaquettes.
This case was already analyzed in section 2 within the approach of 
self-consistent mean field theory. As a check of the path integral approach
we shall show that one is able to reproduce the same results 
from eq.(\ref{lineacritica}).

In the diagonal case eq.(\ref{lineacritica}) becomes
\be \label{casodiag}
1=\alpha \bigg( \frac{\sum_n e^{-\frac{4}{y}(n+\frac{q}{2e})^2} 
\frac{1}{1-4(n+\frac{q}{2e})^2} }
{\sum_{m}e^{-\frac{4}{y}(m+\frac{q}{2e})^2} } \bigg).
\en
Since $n$ is an integer (\ref{casodiag}) is invariant under the 
shift $\frac{q}{2e}\to\frac{q}{2e}+1$.
For $q=0$ eq.(\ref{casodiag}) reduces to (\ref{6}). 
From figure 1
one readily sees that there is no superconductivity for $\alpha<1$.
Due to the periodicity of (\ref{casodiag}) 
this holds for any integer $q$.
For $q/2e$ equal to $1 / 2$ one gets equation (\ref{10}).
From figure 2 one sees that superconductivity is 
attained for all the values of $\alpha$, since the superconducting order 
parameter at zero temperature is different from zero.

For the self-charging model the system exhibits 
superconductivity for all the values of $\alpha$ also if the 
distribution of offset charges is such that integer and 
half-integer charges coexist on the lattice. 
If one denotes by $f_{0}$ the fraction of integer charges and  by 
$f_{\frac{1}{2}}=1-f_0$ the fraction of half-integer charges, 
eq.(\ref{casodiag}) implies that
$$
\alpha=\bigg( f_0 \frac{   \sum_n e^{-\frac{4}{y}n^2} \frac{1}{1-4n^2} }
{\sum_{m}e^{-\frac{4}{y}m^2} }+f_{\frac{1}{2}}\frac{\sum_n 
e^{-\frac{4}{y}(n+\frac{1}{2})^2} \frac{1}{1-4(n+\frac{1}{2})^2} }
{\sum_{m}e^{-\frac{4}{y}(m+\frac{1}{2})^2} } \bigg)^{-1}.
$$
In fig.3 we plot $T_c$ as a function of $\alpha$ 
for several values of $f_{0}$.
\begin{figure}
\begin{center}
\includegraphics[width=8cm,height=8cm]{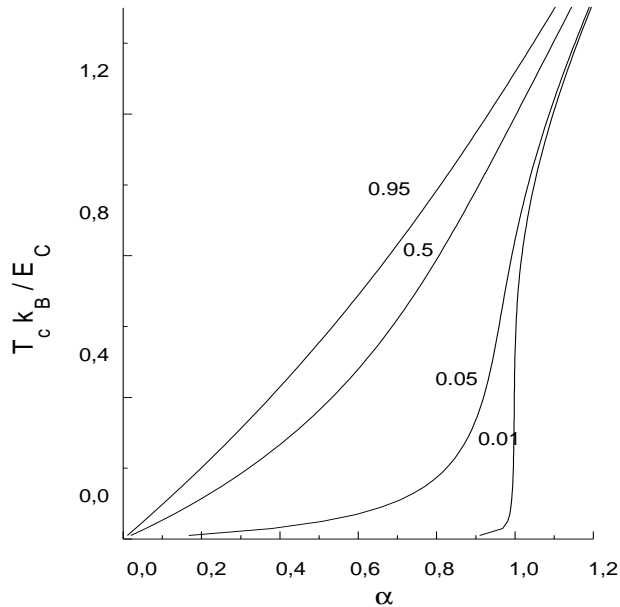}
\end{center}
\caption{Phase diagram for several values of $f_{\frac{1}{2}}$.}
\end{figure}
As expected superconductivity is enhanced as $f_{\frac{1}{2}}$ increases.

\subsection{Models with non-diagonal capacitance matrix}
In \cite{FiSt2} Fishman and Stroud, using a low temperature expansion, 
determined $T_c$ as a function of $\alpha$ for models with non diagonal 
interaction matrix without considering the effect of offset charges.
They did not find signs of normal state reentrance for 
nearest neighbor interaction matrix models 
in which only the diagonal interaction matrix element $U_{\oo\oo}$ 
and the nearest neighbor interaction matrix element $U_{\oo \pp}=
\theta U_{\oo\oo}$ are nonzero.
This can be seen from the expansion of the critical line 
eq.(\ref{lineacritica}) for $q=0$ and small critical temperatures: 
$$
\alpha=1+\big[\frac{8}{3}+2z(1-\frac{1}{1-4\theta^2})\big]
e^{-\frac{4}{y}}+...
$$
Reentrant behavior is possible~\cite{Fa} for $\theta>\theta_c= 
\frac{1}{\sqrt{4+3z}}$ when the coefficient of the exponential 
$e^{-4/y}$ is negative; in fact, the phase boundary line 
$\alpha=\alpha(T_c)$ first bends to the left due  
to the negative coefficient of $e^{-4/y}$ and finally,
when the critical temperature is high enough,
bends to the right, favoring the insulating phase.

As evidenced by Fishman and Stroud~\cite{FiSt2}, the 
regime of physical interest is $\theta <\frac{1}{z}$; namely, when
the capacitance matrix is invertible.
Reentrance is possible only in one dimension ($\theta_c=1 / 
\sqrt{10}< 1 / z=1 / 2$); in higher dimensions reentrance occurs only 
when the electrostatic interaction is long ranged~\cite{FiSt2}.
 
If there are half-integer offset charges on the sites of a square lattice, 
our analysis shows that the equation for the critical line is
\be \label{Efetov}
\alpha=\frac{\sum_{[n_{\ii}]}e^{-\frac{4}{y}\sum_{\ii \jj}
\frac{U_{\ii \jj}}{U_{\oo\oo}}(n_{\ii}+\frac{1}{2})(n_{\jj}+
\frac{1}{2})}}{\sum_{[n_{\ii}]}\frac{e^{-\frac{4}{y}\sum_{\ii \jj} 
\frac{U_{\ii \jj}}{U_{\oo\oo}}(n_{\ii}+\frac{1}{2})(n_{\jj}
+\frac{1}{2})}}{1-
4[\sum_{\jj}\frac{U_{\oo\jj}}{U_{\oo\oo}}(n_{\jj}+\frac{1}{2})]^2}   }.  
\en
In eq.(\ref{Efetov}) appears the expression
\be E[n_{\ii}]= \sum_{\ii\; \jj}
\frac{U_{\ii \jj}}{U_{\bf 00}}({n}_{\ii}+\frac{1}{2})(n_{\jj}+\frac{1}{2})  
\en
which is the electrostatic energy of a generic charge 
distribution on the lattice.

Denoting with $n^0_{\ii}$ and $n^1_{\ii}$ the charge distributions of the 
two lowest lying energy states and with $E^0$ and $E^1$ the 
corresponding energies,
the low temperature expansion of eq.(\ref{Efetov}) 
yields

\be \label{expans}
\alpha=
\frac{ \sum_{[n^0]}e^{-\frac{4}{y}E^0}+\sum_{[n^1]}e^{-\frac{4}{y}E^1}
+\cdots}{   \sum_{[n^0]}
\frac{e^{-\frac{4}{y}E^0}}{1-4\big[\sum_{\jj}
\frac{U_{\oo\jj}}{U_{\oo\oo}}(n_{\jj}^0+\frac{1}{2})\big]^2}
+\sum_{[n^1]}\frac{e^{-\frac{4}{y}E^1}}{1-4\big[\sum_{\jj}
\frac{U_{\oo\jj}}{U_{\oo\oo}}(n_{\jj}^1+\frac{1}{2})\big]^2}+\cdots }\ .
\en

Independently on the explicit form of $U_{\ii\jj}$, 
$E[n_{\ii}]$ reaches its minimum value when 
$(n_{\ii}^0+\frac{1}{2})=\pm\frac{1}{2}(-1)^{i_1+i_2+...+i_D}$ 
with $i_j\ (j=1,...,D)$ 
the components of the lattice position vector $\ii$ in units of the lattice 
spacing.
This charge configuration is exhibited in figure 4.
\begin{figure}[ht]
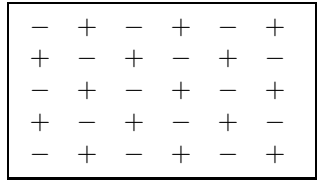

$$ 
\fbox{$ {\begin{array}{cccccc}
-&+&-&+&-&+\\
+&-&+&-&+&-\\
-&+&-&+&-&+\\
+&-&+&-&+&-\\
-&+&-&+&-&+
\end{array} }$ }
$$
\caption{ground state.}
\end{figure}
For models with nearest-neighbor interaction, $i.e.$ $U_{\ii\jj}=
\delta_{\ii\jj}+\theta\sum_{\pp}\delta_{\ii+\pp,\jj}$
with $\sum_{\pp}$ denoting summation over nearest neighbors,
the charge configuration corresponding to the first excited state 
is given in fig.5.
\begin{figure}[ht]
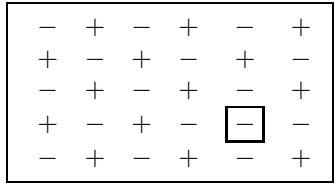

$$ \fbox{ ${\begin{array}{cccccc}
-&+&-&+&-&+\\
+&-&+&-&+&-\\
-&+&-&+&-&+\\
+&-&+&-&\fbox{$-$}&-\\
-&+&-&+&-&+
\end{array}}$} 
$$
\caption{first excited state.}
\end{figure}
The energy of the charge distribution of fig.5 is
$E[n_{\ii}^1]=E[n_{\ii}^0]+z\theta$, where $E[n^0_{\ii}]$, 
the ground state energy, is given by
$\sum_{\ii}\frac{1}{4}(1-z\theta)$.

With the above values of  $E[n_{\ii}^0]$ and $E[n_{\ii}^1]$  
and keeping only 
the leading order term in $T_c$, 
eq.(\ref{expans}) becomes [see Appendix C]
\be \label{moti2} 
\alpha=\bigg(1-(1-z\theta)^2\bigg) \cdot
\Bigg\{1+
\bigg[\bigg(1-\frac{1-(1-z\theta)^2}{1-(1+z\theta)^2}\bigg)
+z\bigg(1-\frac{1-(1-z\theta)^2}{1-(1-(z-2)\theta)^2}\bigg) 
\bigg]e^{-\frac{4}{y}z\theta}+
\cdots \bigg\}\ .
\en 

Reentrant behavior at low temperature occurs when the coefficient 
of the exponential is negative, namely when
\be \label{moti}
a_1 \equiv \bigg(1-\frac{1-(1-z\theta)^2}{1-(1+z\theta)^2}\bigg)
+z\bigg(1-\frac{1-(1-z\theta)^2}{1-(1-(z-2)\theta)^2}\bigg)<0\ .\en
In Appendix C we compute also the coefficients $a_2$ and $a_3$ of the 
higher order exponentials in the expansion (\ref{moti2}).
In fig.6 we plot the coefficients $a_1,\ a_2$ and $a_3$ 
as a function of $\theta$ for 
$z=6$, $i.e.$ for a 3-D array on a square lattice.
\begin{figure}
\begin{center}
\includegraphics[width=8cm,height=8cm]{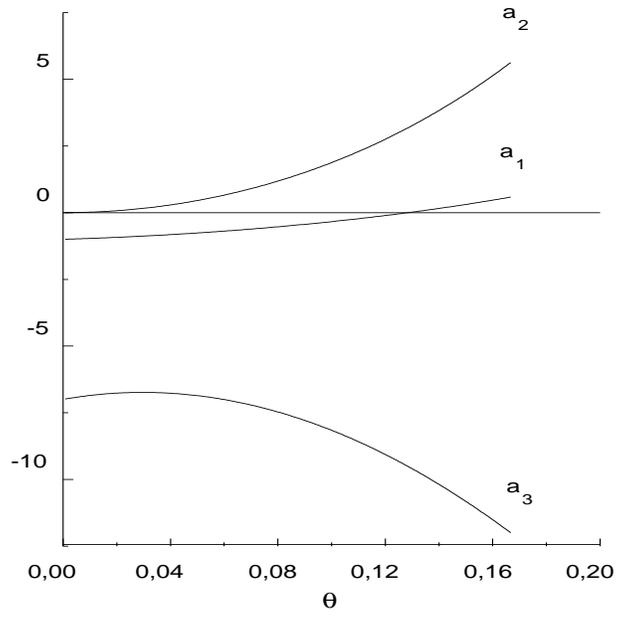}
\end{center}
\caption{Expansion coefficients of $\alpha$ as a function of nearest 
neighbor interaction $\theta$.}
\end{figure}
One sees that the inequality (\ref{moti}) can
be satisfied for values of $\theta$ consistent with the
physical constraint $\theta<1/z=1/6$.\

In fig.7 we plot $T_c$ versus $\alpha$ for $\theta=0.05$ and z=6.
In this plot we keep into account also the next two orders
of (\ref{moti2}) with coefficients $a_2$ and $a_3$.
\begin{figure}
\begin{center}
\includegraphics[width=8cm,height=8cm]{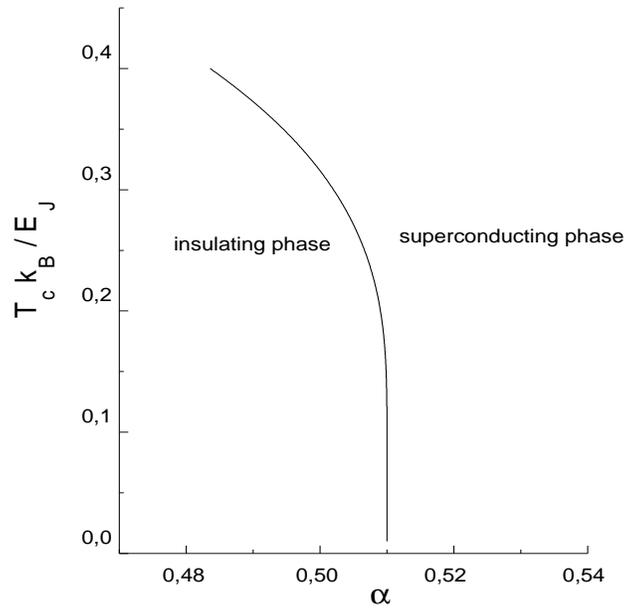}
\end{center}
\caption{Phase diagram for small critical temperatures with $z=6$ 
and $\theta=0.05$.}
\end{figure}
The resulting diagram exhibits reentrance in the insulating phase 
even for models with nearest neighbors interaction.

In fig.8 we plot $\alpha_0=\alpha(T_c=0)$ as a function of $\theta$ 
for $q$ integer and $q$ half-integer and for $z=6$.
\begin{figure}
\begin{center}
\includegraphics[width=8cm,height=8cm]{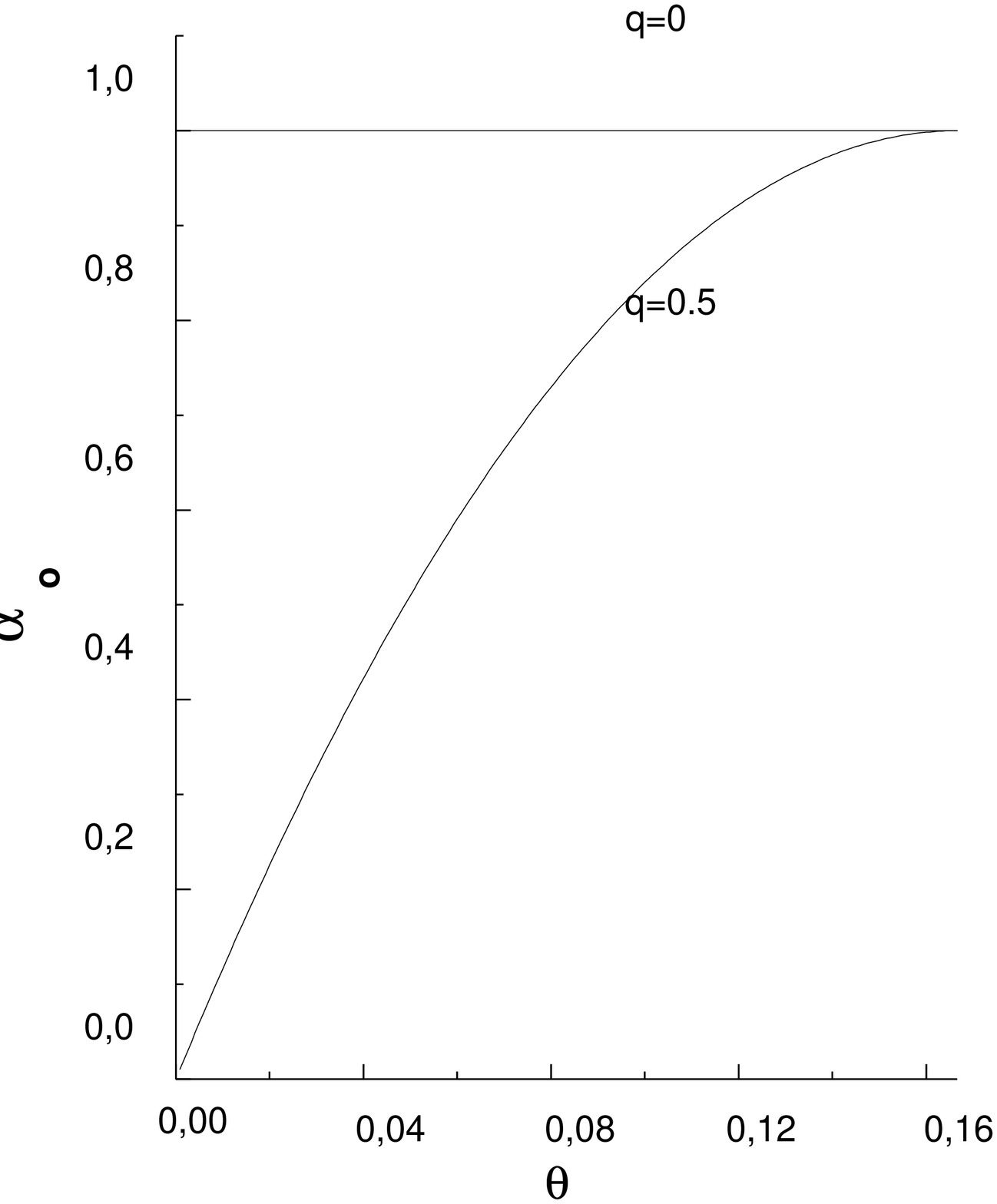}
\end{center}
\caption{Broadening of the superconducting phase at $T=0$ 
with $z=6$ and nearest neighbor interaction.}
\end{figure}
The plot shows that half-integer offset charges always favor
superconductivity and that -at variance with the self-charging model- 
for non-diagonal interaction matrix there is 
always a range of $\alpha$ in which the system is in the insulating phase. 
The plot also shows that for $q/2e=1/2$ and $T=0$ 
the size of the superconducting
region in the phase diagram depends on $\theta$.

\section{Discussion}

In this paper, using the path integral approach, we provided an explicit
derivation of the equation for the phase boundary for quantum
Josephson junction arrays with offset charges and non-diagonal
capacitance matrix. 

For the model with nearest neighbor capacitance
matrix and uniform offset charge $q=1/2$ (in units of $2e$), 
using a procedure developed 
in~\cite{FiSt2}, we were able to determine, 
in the low temperature expansion, the most relevant contributions 
to the equation for the phase boundary.
For this purpose we explicitly constructed the charge distributions
on the lattice corresponding to the lowest energies.

Confirming the results of the numerical analysis of ref.~\cite{FaSh2},
we found a reentrant behavior even with a short ranged interaction.
Our analysis extends the results found in~\cite{FiSt2} to the situation 
in which offset charges are present and provides a physical
picture of the states contributing to the reentrant behavior.

For a model with diagonal capacitance matrix our analysis confirms the
absence of reentrant behavior for the physical situation
where the phase variable is $2\pi$-periodic.
The diagonal model with offset charge $q=1/2$ exhibits 
superconductivity for all the values of $\alpha=z E_J/4E_C$, since 
in this case the superconducting order parameter is different from 
zero at zero temperature; this is evidenced by eq.(\ref{3.35}) in 
Appendix A. An offset charge $q=1/2$ tends to decrease the 
charging energy and thus favors the superconducting behavior 
even for small Josephson energy $E_J$.

A merit of the path integral approach, used in this paper, 
is that it allows to follow at each stage of the analysis the effects 
of the $2\pi$-periodicity of the phase variable. In fact, one can untwist
this periodicity by introducing a set of integers, so that the partition 
function factorizes as a product of a topological term, depending only on 
this set of integers, and a non-topological one explicitly 
evaluated in Appendix B. The Poisson resummation formula for the 
topological part of the partition function turns out very useful for the
derivation of the low critical temperature expansion.

It would be intersting to investigate the superconducting-insulating
behavior in quantum JJA in lower dimensional models, where mean field
theory is not expected to provide accurate results. For $D=1$ there is 
evidence~\cite{larkin} for a new phase separating the superconducting 
and the insulating phase. The analysis of the phase diagram for this case
should be carried out with different methods such as 
the renormalization group~\cite{GC,Jo}.

\appendix
\section{Derivation of the self consistency equation (\ref{10})}
With a uniform charge frustration $q$ the pertinent Mathieu
equation is given by 
\be \label{mat2}
\Big[
-\frac{d^2}{d\varphi^2}-2i \frac{q}{2 e} \frac{d}{d\varphi}+
\bigg(\frac{q}{2 e} \bigg)^2-\alpha <\cos \varphi> \cos \varphi 
\Big]\psi_n(\varphi)=\frac{E_n}{4 E_C} \psi_n(\varphi)\nonumber
\en
Upon defining 
\be \label{3.28}
\psi_n(\varphi)=e^{-i\frac{q}{2e}\varphi} \rho_n(\varphi)
\en
eq.({\ref{mat2}) becomes
\be \label{3.29}
\frac{d^2\rho_n}{d\varphi^2}+
\bigg(\frac{\lambda}{4}-\frac{v}{2}\cos \varphi \bigg)\rho_n =0
\en
with $\lambda_n=E_n /E_C$ and $v=- z E_J 
<\cos \varphi>/2 E_C$. 
Eq.(\ref{3.29}) yields the canonical form of the Mathieu 
equations~\cite{Abramowitz} 
\be
\frac{d^2y}{dx^2}+(\lambda-2 v \cos 2x) y=0\ ,
\label{eqmath}
\en
if one puts $\varphi=2 x$ e $\psi_n=y$.

The Mathieu equation has the well known periodic 
solutions~\cite{Abramowitz}:
$$
\left\{\begin{array}{ll}
            ce_{2n} (x,v) &
\mbox{even solutions with period $\pi$} \\& 
\mbox{with eigenvalues $a_{2n} (v)$}\\
\\se_{2n+2} (x,v) &
\mbox{odd solutions with period $\pi$}\\& \mbox{with eigenvalues 
$b_{2n+2} (v)$}\\
\\ce_{2n+1} (x,v) &\mbox{even solutions with period  
$2\pi$}\\& \mbox{with eigenvalues $a_{2n+1} (v)$}\\
\\se_{2n+1} (x,v) &\mbox{odd solutions with period $2\pi$}\\& 
\mbox{with eigenvalues $b_{2n+1} (v)$}\\
\\n=0,1,2,\ldots 
       \end{array}\ .
\right.
$$ 
If $q/2e$ is integer, the periodic boundary conditions $\psi_n(\varphi=0)=
\psi_n(\varphi=2 \pi)$ singles out only the $2\pi$-priodic 
Mathieu eigenfunctions 
$ce_{2n}$, $se_{2n}$. With these eigenfunctions one may derive 
(\ref{6}) \cite{Sima2}.
If $q/2e$ is half-integer, 
the periodic boundary conditions together with (\ref{3.29}) single out
the $\pi$-anti-periodic Mathieu eigenfunctions 
($i.e.$ $\rho_n$ is anti-periodic of $2\pi$ 
and periodic of $4\pi$). 
These are the Mathieu eigenfunctions $ce_{2n+1}$ and $se_{2n+1}$. 

Since, near the critical temperature $T_c$, the order 
parameter $<\cos \varphi>$ and $v$ are small, 
apart from the phase factor $e^{-i\varphi/2}$ 
(important only for the periodicity), to first order in $v$,
eq.(\ref{mat2}) has the solutions
\be\label{eigenfun}   
\left\{\begin{array}{ll}
\psi_1^e =\frac{1}{\sqrt{\pi}} 
\Bigl( \cos \frac{\varphi}{2} -\frac{v}{8} \cos \frac{3 \varphi}{2} 
\Bigr)\\ \psi_1^o =\frac{1}{\sqrt{\pi}} \Bigl( \sin \frac{\varphi}{2} 
-\frac{v}{8} \sin \frac{3 \varphi}{2} \Bigr)\\ 
            \psi_{2n+1}^e =\frac{1}{\sqrt{\pi}} 
\Bigl \{ \cos \frac{(2n+1)\varphi}{2}-v \Bigl[ 
\frac{\cos \frac{(2n+3)\varphi}{2}}{4 (2n+2)} -
\frac{\cos \frac{(2n-1)\varphi}{2}}{8n} \Bigr] \Bigr \}\\
            \psi_{2n+1}^o =\frac{1}{\sqrt{\pi}} \Bigl \{ 
\sin \frac{(2n+1)\varphi}{2}-v \Bigl[ \frac{\sin \frac{(2n+3)
\varphi}{2}}{4 (2n+2)} -\frac{\sin \frac{(2n-1)\varphi}{2}}{8n} \Bigr] 
\Bigr \}\\(n=1,2,\ldots)
\end{array}
\right.
\en
with the corresponding eigenvalues given by
\be \label{3.34}
\left\{\begin{array}{ll}E_1^e=E_C (1+q)\\
\\ E_1^o=E_C (1-q)\\ \\
E_{2n+1}^e=E_{2n+1}^o=E_C (2n+1)^2\\\\ (n=1,2,\ldots)
\end{array}\ .
\right.
\en
The expectation values of the superconducting order parameter on the
eigenfunctions (\ref{eigenfun})
 are given by
\be
<\psi_n|\cos\varphi|\psi_n>=
\int\limits_{0}^{2 \pi} d\varphi \cos\varphi \mid 
\psi_n (\varphi) \mid ^2\ .
\en
Using (\ref{eigenfun}), to the first order in $v$ one gets
\be
\left\{\begin{array}{ll}
<\psi_1^e|\cos \varphi|\psi_1^e>={1 \over 2}-\frac{v}{8}\\          
<\psi_1^o|\cos \varphi|\psi_1^o>=-{1 \over 2}-\frac{v}{8}\\ 
<\psi_{2n+1}^e|\cos \varphi|\psi_{2n+1}^e>=\frac{v}{8n(n+1)}\\
<\psi_{2n+1}^o|\cos \varphi|\psi_{2n+1}^o>=\frac{v}{8n(n+1)}\\   
(n=1,2,\ldots)
\end{array}\ .
\right.
\label{3.35}
\en

Inserting (\ref{3.34}) and (\ref{3.35}) in (\ref{5}) and
keeping only the terms proportional to $v \sim <\cos \varphi>$, one finds 
\be
1=\alpha \> \> \frac{(\frac{2}{y}+\frac{1}{2}) e^{-\frac{1}{y}}-
\sum\limits_{n=1}^{\infty} \frac{e^{-\frac{1}{y}(2n+1)^2 }}{2n(n+1)}}
{2e^{-\frac{1}{y}}+2\sum\limits_{n=1}^{\infty} e^{-\frac{1}{y}(2n+1)^2}}\,
\label{3.36}\ ;
\en
namely eq.(\ref{10}).

\section{The phase correlator}
In this appendix we want to elucidate the
computation of the correlator defined in equation (\ref{21}).
For this purpose one should compute the path integral
\be \label{appcor}
\frac{\int D \phi_{\ii} e^{i\phi_{\bf r}(\tau)-i\phi_{\bf s}(\tau')} 
\exp \grs \intau (-\frac{1}{2}C_{\bf ij}
\frac{\dot{\phii}}{2e}\frac{\dot{\phij}}{2e})\grd}  
{\int D \phi_{\ii} \exp \grs \intau (-\frac{1}{2}C_{\bf ij}
\frac{\dot{\phii}}{2e}\frac{\dot{\phij}}{2e})\grd}\ .
\en
Fourier transforming $\phii(\tau)$ according to
\be
\phii(\tau)=\frac{1}{\beta}\sum_{n=-\infty}^{+\infty}
\phi_{\ii, m}e^{i \omega_{m}\tau}
\en
with $0 \le \tau \le \beta$ and $\omega_m=\frac{2\pi}{\beta}m$,
the numerator of (\ref{appcor}) becomes
$$ 
\int\prod_{\ii}d\phi_{\ii,0} 
\prod _{n=1}^{\infty}d\phi_{\ii,n}d\phi_{\ii,n}^*
\exp \bigg\{-\frac{1}{4e^2\beta}\sum_{\ii \jj}\sum_{n=1}^{+\infty}
C_{\ii \jj}\omega_n^2\phi_{\ii,n} \phi_{\jj,n}^*+$$
\be \label{intermed}
+\frac{i}{\beta}\sum_{n=1}^{\infty}\left(\phi_{\rr,n}e^{i \omega_n \tau}-
\phi_{\es,n}^*e^{-i \omega_n \tau'}\right)+c.c.
+ \frac{i}{\beta}\left(\phi_{\rr,0}-\phi_{\es,0}\right) \bigg\}\ .
\en

Upon 
integrating over the components $\phi_{\rr,0}$, $\phi_{\es 0}$ 
one gets a factor $\delta_{\rr \es}$ 
\be\label{Kappa}
\bigg(\prod_{\ii\ne\rr,\es}\int_{-\infty}^{\infty}d\phi_{\ii,0}\bigg)
\bigg(\int_{-\infty}^{\infty}d\phi_{\rr,0}\int_{-\infty}^{\infty}d
\phi_{\es,0}e^{\frac{i}{\beta}(\phi_{\rr,0}-\phi_{\es,0})}\bigg)=
\delta_{\rr \es} \cdot K
\en
where $K$ is an irrelevant divergent constant which cancels 
against the denominator.
Using (\ref{Kappa}), (\ref{intermed}) becomes
$$
K \delta_{\rr \es}\prod_{n=1}^{\infty}
\int_{-\infty}^{\infty}\prod _{\ii}d\phi_{\ii,n}d\phi_{\ii,n}^*
\exp\left(-\frac{1}{4e^2\beta}\sum_{\ii \jj}
C_{\ii \jj}\omega_n^2\phi_{\ii,n} \phi^*_{\jj,n}
\right. \>\> \cdot $$ $$ \cdot \> +\left.\sum_{\ii}\phi_{\ii,n}
\frac{i}{\beta}\delta_{\rr\ii}(e^{i \omega_n \tau}-
e^{i \omega_n \tau'})-\sum_{\ii}\phi^*_{\ii,n}
\delta_{\rr \ii}\frac{i}{\beta}
(e^{-i \omega_n \tau'}-e^{-i \omega_n \tau'})\right)\ .
$$
The  multiple Gaussian integral may be easily computed to give,
up to an irrelevant constant which cancels against the denominator, 
$$
\delta_{\rr \es}\prod_{\ii}\prod_{n=1}^{\infty}\int_{-\infty}^{\infty} 
d\phi_{\ii n} d\phi^*_{\ii n}\exp \bigg\{\sum_{\ii\jj}
\frac{i}{\beta}\delta_{\rr\ii}(e^{i \omega_n \tau}-
e^{i \omega_n \tau'}) \cdot$$
$$\cdot \big(\frac{4e^2\beta C_{\ii \jj}^{-1}}{\omega_n^2}\big)
\frac{i}{\beta}\delta_{\rr\ii}(e^{-i \omega_n \tau}-
e^{-i \omega_n \tau'})\bigg\}
=$$
$$
= \delta_{\rr \es} \exp\bigg\{\frac{8 e^2C_{\rr \rr}^{-1}}{\beta }
\sum_{n=1}^{\infty}(\frac{1-\cos 
\omega_n(\tau-\tau' )}{\omega_n^2}) \bigg\}=
$$
\bea \nonumber
=\delta_{\rr \es}\exp \grs-2e^2C^{-1}_{\rr \rr}
\biggl(|\tau-\tau'|-\frac{(\tau-\tau')^2}{\beta} \biggr)\grd 
\ena
where $ -\beta \le \tau-\tau'
\le \beta$.
In the last step, the identity
\be
|x|-\frac{x^2}{\beta}=\sum_{n=1}^{\infty}(\frac{4}{\beta \omega_n^2}-
\frac{4\cos \omega_n x}{\beta \omega_n^2}) \hspace{1cm}
-\beta \le x \le \beta
\en
has been used.
This completes the proof of (\ref{22})

\section{Low $T_c$ expansion}

In this appendix we derive equation (\ref{moti2}) and compute 
the next two orders whose coefficients are plotted in fig.6.
Using the notation $(-1)^{\ii}=(-1)^{i_1+\cdots+i_D}$,
the ground state charge configuration $n_{\ii}^0$ can be written as 
$$ (n_{\ii}^{0}+\frac{1}{2})=\frac{1}{2}(-1)^{\ii}\ .$$
The first excited states read 
$$(n_{\ii}^{1_{\rr}}+\frac{1}{2})=n_{\ii}^{0}(1-2 \delta_{\rr \ii})\ ;$$
where the apex $1_{\rr}$ means that this first excited state is 
obtained from the ground state by flipping the sign of the charge at the
site $\rr$. Higher excitations may be obtained 
from the ground state by flipping the sign of two charges at sites 
$\rr$ and $\es$ and can be represented as
$$(n_{\ii}^{2_{\rr\es}}+\frac{1}{2})=n_{\ii}^{0}(1-2 
\delta_{\rr \ii}-2 \delta_{\es \ii})\ ,$$

The energy shifts are given by
$$\Delta^{1}=E^1-E^0=\sum_{\ii \atop \ii\ne \rr}
U_{\ii \rr}(-1)^{\rr+\ii+1} $$ 
and
$$\Delta^{2_{\rr\es}}=  E[n_{\ii}^{2_{\rr\es}}]-E^0= 
2\Delta^{1}+2(-1)^{\rr-\es}U_{\rr\es}\ .$$
Note that, whereas the energy $E^1$ of the charge configurations
$n_{\ii}^{1_{\rr}}$ does not depend on $\rr$, $E[n_{\ii}^{2_{\rr\es}}]$
depends on the relative position $\rr-\es$
of the charges whose sign has been flipped.

Defining
$$
R^0=\frac{1}{1-4[\sum_{\jj}\;
U_{\oo\jj}(n_{\jj}^{0\pm}+\frac{1}{2})]^2}\ ,
$$
$$R^{1_{\rr}}=\frac{1}{1-4[\sum_{\jj}\;
U_{\oo\jj}(n_{\jj}^{1_{\rr}}+\frac{1}{2})]^2}
$$
and
$$
R^{2_{\rr\es}}=\frac{1}{1-4[\sum_{\jj}\;
U_{\oo\jj}(n_{\jj}^{2_{\rr\es}}+\frac{1}{2})]^2},
$$
one may expand eq.(\ref{Efetov}) for small critical 
temperatures $(y \propto T_c \to 0)$, according to
$$
\alpha=\frac{1+\sum_{\rr} e^{-\frac{4}{y}\Delta^1}+ \sum^*_{\rr \ne\es}
e^{-\frac{4}{y}\Delta^{2_{\rr \es}}}+\cdots  }{  R^0+\sum_{\rr}R^{1_{\rr}} 
e^{-\frac{4}{y}\Delta^1}+ \sum^*_{\rr \ne\es}R^{2_{\rr \es}}e^{-\frac{4}{y}
\Delta^{2_{\rr \es}}}+\cdots}
=$$
\be \label{svi}
=\frac{1}{R^0}\bigg[1+\sum_{\rr} (1-\frac{ R^{1_{\rr}} }{R^0})  
e^{-\frac{4}{y}\Delta^1}+ \sum^*_{\rr \ne\es} (1-\frac{ R^{2_{\rr \es}} }{R^0})
e^{-\frac{4}{y}\Delta^{2_{\rr \es}}}+$$
$$\sum_{\rr \es}( \frac{ R^{1_{\rr}} }{R^0}\frac{ R^{1_{\es}} }{R^0}
-\frac{ R^{1_{\rr}} }{R^0}) e^{-\frac{8}{y}\Delta^1}+\cdots\bigg]
\en
where $\sum^*_{\rr\ne \es}$ indicates a summation over pairs 
of different sites $\rr, \es$, where each pair is counted only once.

For a nearest neighbor interaction
$U_{\oo\jj}=\delta_{\oo\jj}+\theta \sum_{\pp}\delta_{\jj\pp}$ (where $\pp$
denotes the vector connecting two neighboring sites)
one has
$$
\Delta^1=z\theta,    
$$
$$
\Delta^{2_{\rr\es}}=
\left\{  \begin{array}{lcl}
2(z-1)\theta & &\mbox{ $\rr-\es =\pp$  }\\
   2 z \theta && \mbox{ $\rr-\es \ne \pp$ }
\end{array} \right. 
$$
$$
R^0=\frac{1}{1-(1-z \theta)^2}
$$
\[ R^{1_{\rr}}=\left\{ \begin{array}{ccc}  
  R^0 &&
\mbox{  $\rr \ne \oo, \pp $}\\  
 \frac{1}{1-(1+z\theta)^2} && \mbox{ $\rr=\oo$} \\ 
\frac{1}{1-(1-(z-2)\theta))^2} &&\mbox{ $\rr=\pp$ }
\end{array}\right.  \]
\[
R^{2_{\rr\es}}=\left\{ \begin{array}{ccc} R^0 && \mbox{ 
$\rr,\es \ne \oo,\pp$ } \\
  R^{1_{\es}} &&\mbox{ $\rr \ne \oo,\pp$}\\
\frac{1}{1-(1+(z-2)\theta)^2}&&\mbox{ $\rr=\oo$  $\es=\pp$ } \\
 \frac{1}{1-(1-(z-4)\theta)^2}&&\mbox{ $\rr=\pp$ $\es=\pp'$} \end{array} 
\right. 
\]
Substituting these relations in (\ref{svi}),
one obtains the expansion for small temperatures
of the critical line equation, up to the first four orders 
\be\label{a1a2a3}
\alpha=\bigg(1-(1-z\theta)^2\bigg)\times\bigg(
1+a_1\>e^{-\frac{4}{y}z\theta}+a_2\>e^{-\frac{8}{y}(z-1)\theta}+
a_3\>e^{-\frac{8}{y}z\theta}\bigg)\ .
\en
$a_1$ is given in (\ref{moti}),
$a_2$ is equal to
$$
(z-1)z\bigg(1-\frac{1-(1-z\theta)^2}{1-(1-(z-2)\theta)^2}\bigg)+
z\bigg(1-\frac{1-(1-z\theta)^2}{1-(1+(z-2) \theta)^2}\bigg) $$ 
and $a_3$ is given by 
$$
\bigg(\frac{1-(1-z\theta)^2}{1-(1+z\theta)^2}
\bigg)^2-\bigg( \frac{1-(1-z\theta)^2}{1-(1+z\theta)^2}\bigg) 
+z(z-1) \cdot \bigg( \frac{1-(1-z\theta)^2}{1-(1-(z-2)\theta)^2}-1 \bigg)+$$
$$+z^2 \cdot \frac{1-(1-z\theta)^2}{1-(1-(z-2)\theta)^2}
\bigg(\frac{1-(1-z\theta)^2}{1-(1-(z-2)\theta)^2}-1 \bigg)+z 
\cdot \frac{1-(1-z\theta)^2}{1-(1+z\theta)^2} 
\bigg( \frac{1-(1-z\theta)^2}{1-(1-(z-2)\theta)^2}-1 \bigg)+$$
$$+z \cdot \frac{1-(1-z\theta)^2}{1-(1-(z-2)\theta)^2)}\cdot
\bigg(\frac{1-(1-z\theta)^2}{1-(1+z\theta)^2}-1  \bigg)+
\frac{z (z-1)}{2}\cdot \bigg( 1- 
\frac{1-(1-z\theta)^2}{1-(1-(z-4)\theta)^2} \bigg)\ .$$ 
The condition  for the reentrant behavior is $a_1<0$. 
In fig.6 we plot the coefficients $a_1$, $a_2$, $a_3$ as a function of 
$\theta$.  
In fig.7 we plot the critical equation (\ref{a1a2a3}) with $\theta=0.05$
and $z=6$.              

\vskip 0.3truein  
{\large \bf Acknowledgements:}
G. G. and P. S. are happy to acknowledge the many 
discussions with their friend G. W. Semenoff who joined their
efforts in the initial stage of this work.
We thank R. Fazio, M. Rasetti and A. Tagliacozzo for discussions.
The research is supported by a grant from the theory group of I.N.F.M. 
and by M.U.R.S.T. .


\begin{thebibliography}{bibliografia}

\bibitem{sima} E. Simanek, Inhomogeneous Superconductors, 
Oxford University Press,
(1994).
\bibitem{Joseph}B. D. Josephson, Phys. Lett. {\bf 1},7 (1962).
\bibitem{And}P.W. Anderson,in 
{\em Lectures on the Many Body Problem}, edited by E.R. 
Caianello (Academic, New York, 1964), p.113.
\bibitem{Ab}B. Abeles, Phys. Rev. B {\bf 15}, 2828 (1977).
\bibitem{Ef}K. B. Efetov, Zh. Eksp. Teor. Fiz. {\bf 78}, 2017 (1980) 
[Sov. Phys. JETP {\bf51}, 1015 (1980)]. 
\bibitem{larkin}A. I. Larkin and L. I. Glazman, cond-mat/9705169.
\bibitem{FaSh1}C. Bruder, R. Fazio, A. Kampf, A. van Otterlo and 
G. Sch\"{o}n, Phys. Scri. {\bf42}, 159 (1992).
\bibitem{FaSh2}A. van Otterlo, K. H. Wagenblast, R. Fazio and 
G. Sch\"{o}n, Phys. Rev. B {\bf 48}, 3316 (1993). 
\bibitem{Sima1}E. Sim\`anek, Solid State Commun. {\bf 31}, 419 (1979).
\bibitem{Sima2}E. Sim\`anek, Phys. Rev. B
{\bf 22}, 459 (1980).
\bibitem{Sima3}E. Si\`manek, Phys. Rev. B {\bf 23}, 5762 (1981).
\bibitem{Fa}P. Fazekas, Z. Phys. B, {\bf 45}, 215 (1982).
\bibitem{Kiss}J. G. Kissner  and U. Eckern, Z. Phys. {\bf B91}, 155 (1993).
\bibitem{Simkin}M. V. Simkin, Phys. Rev. B {\bf 44}, 7074  (1991).
\bibitem{GC}E. Granato and M. A. Continentino, 
Phys. Rev. B {\bf 48}, 15 977 (1993).
\bibitem{Jo} C. Rojas and J. V. Jos\'e, cond-mat/9610051 .
\bibitem{Fa1}P. Fazekas, B. Muhlschlegel and M. Schroter, Z. Phys. 
{\bf B57}, 193 (1984).
\bibitem{FiSt1}R.S. Fishman and Stroud, Phys. Rev. B {\bf 35}, 1676 (1987).
\bibitem{Don}S. Doniach. Phys. Rev. B {\bf 24}, 5063 (1981).
\bibitem{FiSt2}R.S. Fishman and Stroud, Phys. Rev. B {\bf 37}, 1499 (1988). 
\bibitem{Fi} R. S. Fishman, Phys. Rev. B {\bf 42} 1985 (1990).
\bibitem{RS}E. Roddick and D. Stroud, Phys. Rev. B {\bf 48}, 16 600 (1993).
\bibitem{Luc}G. Luciano, U. Eckern and J.G. Kissner, Europhys. Lett. 
{\bf 32}, 8 (1995).
\bibitem{hub}J. Hubbard, Phys. Rev. Lett. {\bf 3}, 77 (1959);
R. L. Stratonovich, Sov. Phys. Dokl. {\bf 2}, 416 (1958).
\bibitem{Abramowitz} M.Abramowitz and I.A.Stegun, 
{\em Handbook of Mathematical Functions},  Dover, New York, 1964.
\bibitem{schon}G. Schon and A. D. Zaikin, Physica, {\bf 152 B}, 203 (1988).
\bibitem{likh} K. K. Likharev and  A. B. Zorin, J. Low. Temp. Phys. 
{\bf 59}, 347 (1985).
\bibitem{Simkin1}M. V. Simkin, cond-mat/9607001.
\bibitem{Sima4}E. Simanek, Phys. Rev. B {\bf 32} 500 (1985).







\end{thebibliography}
\end{document}